\documentclass[conference]{IEEEtran}
\IEEEoverridecommandlockouts

\usepackage{amsmath,amssymb,amsfonts}   
\usepackage{mathtools}                  
\usepackage{threeparttable}
\usepackage{placeins}
\usepackage{titlesec}

\usepackage{cite}
\usepackage{amsmath,amssymb,amsfonts}
\usepackage[compatibility=false]{caption}
\usepackage{graphicx}
\usepackage{textcomp}
\usepackage{xcolor}
\def\BibTeX{{\rm B\kern-.05em{\sc i\kern-.025em b}\kern-.08em
    T\kern-.1667em\lower.7ex\hbox{E}\kern-.125emX}}

\usepackage{booktabs}                   
\usepackage{url}                        
\usepackage{caption}                    
\usepackage{mwe}                        
\usepackage{threeparttable}
\usepackage{siunitx}

\usepackage{algorithm}                 
\usepackage{algpseudocode}             

\usepackage{xcolor}                    
\usepackage{subfigure}                 
\usepackage{array}                     

\begin{document}

\title{System-Level Comparison of Multimodal and In‑Band mmWave Sensing for Beam Prediction in 6G ISAC}

\author{\IEEEauthorblockN{
Abidemi Orimogunje\thanks{Corresponding author: remero@hanyang.ac.kr. Co-Corresponding author: oigbafe2@hanyang.ac.kr}\IEEEauthorrefmark{1}\IEEEauthorrefmark{2},   
Hyunwoo Park\IEEEauthorrefmark{2},   
Igbafe Orikumhi\IEEEauthorrefmark{2},    
Sunwoo Kim\IEEEauthorrefmark{2},    
Dejan Vukobratovic\IEEEauthorrefmark{3}      
}                                     
\IEEEauthorblockA{\IEEEauthorrefmark{1} 
African Center of Excellence in Internet of Things, University of Rwanda, Rwanda}
\IEEEauthorblockA{\IEEEauthorrefmark{2}
Department of Electronic Engineering, Hanyang University, South Korea}
\IEEEauthorblockA{\IEEEauthorrefmark{3}
Faculty of Technical Sciences, University of Novi Sad, Serbia.}
}
 \maketitle

\begin{abstract}

Integrated sensing and communication (ISAC) can reduce beam‑training overhead in mmWave vehicle-to-infrastructure (V2I) links by enabling in-band sensing-based beam prediction, while exteroceptive sensors can further enhance the prediction accuracy. This work develop a system‑level framework that evaluates camera, LiDAR, radar, GPS, and in‑band mmWave power, both individually and in multimodal fusion using the DeepSense‑6G Scenario‑33 dataset. A latency‑aware neural network composed of lightweight convolutional (CNN) and multilayer‑perceptron (MLP) encoders predict a 64‑beam index. We assess performance using Top‑k accuracy alongside spectral‑efficiency (SE) gap, signal‑to‑noise‑ratio (SNR) gap, rate loss, and end‑to‑end latency.
Results show that the mmWave power vector is a strong standalone predictor, and fusing exteroceptive sensors with it preserves high performance: mmWave alone and mmWave+LiDAR/GPS/Radar achieve 98\% Top-5 accuracy, while mmWave+camera achieves 94\% Top-5 accuracy. The proposed framework establishes calibrated baselines for 6G ISAC-assisted beam prediction in V2I systems. 
\end{abstract}


\section{Introduction}
Future sixth-generation (6G) systems are converging toward integrated sensing and communication (ISAC), where co-located sensors at the infrastructure continuously perceive the environment while sustaining high-rate links \cite{10920855}. At mmWave/THz frequencies, narrow analog beams are essential for overcoming severe path loss and maintaining high signal-to-noise ratio (SNR) \cite{9530500,10438962, 9745525}. However, exhaustive beam sweeping and frequent retraining of such beams introduce latency and control overhead, particularly in vehicle-to-infrastructure (V2I) settings with fast dynamics and frequent line-of-sight (LoS)/non-LoS transitions \cite{10633859, 9869437}. Leveraging exogenous sensing (from camera, radar etc.) to predict beams can reduce in-band training and accelerate link recovery, a direction validated on the real-world DeepSense‑6G dataset and emerging ISAC prototypes \cite{11133431, m2beamllm, 10561505}.

Recent studies demonstrate that (i) vision-position fusion can reach high Top‑3 accuracies with real mmWave data, significantly cutting beam-training overhead \cite{9771835}, (ii) position-only models with commercial Global Positioning System (GPS) achieve modest Top‑1 but still deliver low power loss (rate loss) relative to the optimal beam, suggesting that accuracy alone is not a faithful system metric \cite{10278998}, (iii) improved modality encoders and pooler-style transformers can boost accuracy and reduce complexity \cite{11142617}, and (iv) geometry-aware pipelines using Light Detection and Ranging (LiDAR), and camera for 3D localization can directly steer beams and reduce search overhead \cite{10757880}, with digital-twin ray tracing emerging to make beam bursts obstacle-aware in real time \cite{11174444}.

Building on these insights, we present a complete, validated, and reproducible ISAC-assisted multimodal framework for beam prediction using DeepSense‑6G Scenario‑33 dataset. Our contributions are summarized as follows:
\begin{enumerate}
    \item We compares individual sensing modalities (camera, LiDAR, radar, GPS and in-band mmWave power), their pairwise and multimodal fusions using not only Top‑$k$ accuracy but also spectral efficiency (SE), SNR, rate loss and end-to-end latency, thereby aligning model performance with actual communication behavior.
    \item We treat the 64-dimensional mmWave receive power vector as a first-class ISAC sensing modality alongside camera, LiDAR, radar and GPS. Each sensor is encoded using compact CNN or MLP backbones and a transformer-based fusion module aggregates the embeddings to predict the receive-beam index.
    \item On Scenario‑33, we show that the mmWave power vector alone is a strong anchor modality, while fusing GPS or LiDAR provides consistent gains and camera/radar offer situational improvements. We quantify how small variations in Top-1 accuracy translate into substantial changes in rate loss and SNR gap.
\end{enumerate}
\noindent Collectively, this work advances ISAC-assisted beam prediction by moving beyond accuracy-only comparisons and coupling beam-selection quality with SNR/SE performance and sensing-to-beamforming latency on real 60‑GHz V2I measurements. The framework offers practical guidance and a reproducible baseline for future 6G ISAC deployment.

The rest of the paper is organized as follows. Section~\ref{sec:sys_mod} details the system model. Section~\ref{sec:prop_meth} describes the proposed multimodal architecture, fusion strategy, and training pipeline. Section~IV presents experimental results and Section~V concludes the paper.

\section{System Model}\label{sec:sys_mod}
We consider a V2I ISAC setup with two units\cite{10144504}. \textbf{Unit~1} (infrastructure) is a stationary receiver operating at 60\,GHz, equipped with a \(N_r{=}16\)-element phased array and a receive codebook of size \(B{=}64\). Co‑located sensors; an RGB camera, a 3D LiDAR, a frequency modulated continuous wave (FMCW) radar, and a GPS‑RTK receiver observe the scene. \textbf{Unit~2} (mobile) is a transmitter with a single active antenna element. All sensors are time‑aligned to the mmWave sweep at timestamp \(t\).

\subsection{Communication Model}
At time \(t\), the transmitter emits a complex symbol \(x_t\in\mathbb{C}\). The \(N_r\times 1\) array observation at Unit~1 is

\begin{equation}
  \mathbf{y}_t \;=\; \mathbf{h}_t\, x_t \;+\; \mathbf{n}_t,
\end{equation}
where $\mathbf{h}_t\!\in\!\mathbb{C}^{N_r}$ is the downlink channel seen at the receiver and $\mathbf{n}_t\!\sim\!\mathcal{CN}(\mathbf{0},\sigma^2\mathbf{I})$ is thermal noise. Applying the $i$th receive beam yields the scalar post-combiner output
\begin{equation}
  z_{t,i} \;=\; \mathbf{w}_i^{\!H}\mathbf{y}_t,
\end{equation}
and the corresponding received power measurement
\begin{equation}
 p_{t,i} \;=\; |z_{t,i}|^{2}, \qquad i=1,\dots,64.
\end{equation}
Collecting all beams produces the power vector, $\mathbf{p}_t{=}[p_{t,1},\ldots,p_{t,64}]^\top\!\in\!\mathbb{R}^{64}_{\ge 0}$, which we treat as both a communication measurement and a sensing feature. The optimal beam index at time $t$ is
\begin{equation}
\label{eq:opbeam}
  b_t^\star \;=\; \arg\max_{i \in \{1,\ldots,64\}} p_{t,i}.
\end{equation}
With noise variance $\sigma^2$, the instantaneous post-beamforming SNR and rate for a candidate beam $i$ are
\begin{equation}
  \mathrm{SNR}_{t,i} \;=\; \frac{p_{t,i}}{\sigma^2},
\end{equation}
\begin{equation}
      R_{t,i} \;=\; \log_2\!\big(1+\mathrm{SNR}_{t,i}\big),
\end{equation}
and the oracle spectral efficiency is $R_{t,b_t^\star}$. Throughout, we adopt the standard codebook-based beam selection abstraction used in multimodal beam prediction works.
For a predicted index $\hat b_t$, the achievable rate is
\begin{equation}
  R_{t,\hat b_t} = \log_2\!\left(1+\frac{p_{t,\hat b_t}}{\sigma^2}\right).  
\end{equation}
We define performance gaps that align model quality with system impact:
\begin{align}
{\rm SNR\_gap}_t~[{\rm dB}] &= 10\log_{10}\!\Big(\tfrac{\mathrm{SNR}_{t,b_t^\star}}{\mathrm{SNR}_{t,\hat b_t}}\Big)
 \;
\end{align}
\begin{align}
    {\rm Rate\_loss}_t &= R_{t,b_t^\star} - R_{t,\hat b_t},
\end{align}
\begin{align}
{\rm SE\_opt} &= \mathbb{E}_t[R_{t,b_t^\star}],
\end{align}
\begin{align}
  {\rm SE\_pred} &= \mathbb{E}_t[R_{t,\hat b_t}]  
\end{align}
and additionally report the average power-loss metric advocated for real-world assessment. Oracle SE is always an upper bound, i.e., ${\rm SE\_opt}\ge {\rm SE\_pred}$.

\subsection{Sensing Model}
At every synchronized timestamp, Unit~1 provides the fused sensing set
\begin{equation}
  \mathcal{S}_t \;=\; \big\{\,\mathbf{I}_t,\;L_t,\;\mathcal{R}_t,\;\mathbf{g}_t\},
\end{equation}
where $\mathbf{I}_t,\,L_t,\,\mathcal{R}_t,\,\mathbf{g}_t,$ represent the camera image, point cloud from LiDAR, radar features,and GPS-based features respectively. Each element is summarized as follows.
\subsubsection{RGB Camera}An RGB camera is co-located with the mmWave array, and we use the RGB stream $\mathbf{I}_t\!\in\!\mathbb{R}^{540\times 960\times 3}$ sampled at 30\,fps. It provides semantic and geometric cues within the base station’s field of view.

\subsubsection{LiDAR}
A rotating 3D LiDAR (100\,m range, up to 20\,Hz) returns a point cloud $L_t{=}\{\mathbf{q}_\ell\}_{\ell=1}^{N_L}\!\subset\!\mathbb{R}^3$ ($N_L$ is the number of points in the LiDAR sweep) sampling the environment with approximately uniform azimuth-elevation spacing. LiDAR contributes metrically accurate ranges to structures and moving obstacles, complementing the mmWave power vector by revealing potential blockages and reflecting surfaces that shape the multipath. 

\subsubsection{FMCW Radar}
Each radar frame consists of complex samples with dimensions $(A{\times}S{\times}C)$ corresponding to the number of receiver antennas $A$, samples per chirp $S$, and chirps per frame $C$. After standard preprocessing, we form magnitude range-velocity ($\mathbf{H}_{\mathrm{RV}}$) and range-angle ($\mathbf{H}_{\mathrm{RA}}$) representations via 2D FFTs
and use $\mathcal{R}_t{=}\{\mathbf{H}_{\mathrm{RA}},\mathbf{H}_{\mathrm{RV}}\}$ as radar features. Radar contributes Doppler and angular information that is resilient to illumination in weather conditions and complements camera-LiDAR for dynamic targets.
\subsubsection{GPS-RTK}
A GPS receiver provides a geodetic feature vector $\mathbf{g}_t$ . In position-aided beam management\cite{10278998}, relative geometry derived from GPS can reduce beam training overhead when fused with visual and radar cues, though the raw GPS may exhibit non-Gaussian errors that impact direct beam classification. 
\subsubsection{mmWave Receive Power Vector}
The 64\text{-}dimensional mmWave receive-power vector ($\mathbf{p}_t$) is included as an explicit sensing channel. It captures the instantaneous angular
receive-power profile over the 64‑beam codebook and is therefore highly
informative for beam alignment in learned ISAC systems. Using $\mathbf{p}_t$ as
an input feature is consistent with recent multimodal formulations that fuse
exteroceptive sensing with in-band radio signatures
\cite{10706812}.


\subsection{Problem Formulation}
Let $\mathcal{B}=\{1,\ldots,64\}$ denote the receive-beam codebook and $\mathcal{S}_t$, the synchronized ISAC observation  at time $t$.
The oracle beam index is defined from the radio measurement as (\ref{eq:opbeam})

The goal is to learn a predictor $\mathcal{M}_\theta$ that maps multi-modal observations to a beam posterior $\boldsymbol{\pi}_t=\mathcal{M}_\theta(\mathcal{S}_t; \mathbf{p}_t)\in\mathbb{R}^{|\mathcal{B}|}$ and outputs
\begin{equation}
    \hat b_t \;=\; \arg\max_{i\in\mathcal{B}}~\pi_{t,i}.
\end{equation}
Beyond top-$k$ classification, the problem is system-centric: we seek predictions that (i) maximize spectral efficiency,
\begin{equation}
   \mathrm{SE}_{\text{pred}} \;=\; \mathbb{E}\!\left[\log_2\!\big(1+\tfrac{p_{t,\hat b_t}}{\sigma^2}\big)\right], 
\end{equation}
and (ii) minimize the SNR gap to the oracle,
\begin{equation}
    \mathrm{SNR\_gap} \;=\; \mathbb{E}\!\left[10\log_{10}\!\Big(\tfrac{p_{t,b_t^\star}}{p_{t,\hat b_t}}\Big)\right],
\end{equation}
while maintaining high Top-K accuracy and low sensing-plus-inference latency. Formally, we learn $\theta$ to optimize a multi-objective criterion
\begin{equation}
\min_{\theta}\; \mathbb{E}\Big[\underbrace{\mathrm{CE}(b_t^\star,\boldsymbol{\pi}_t)}_{\text{beam classification}} 
\;+\; \lambda_{\text{gap}}\!\cdot\!\mathrm{SNR\_gap} 
\;+\; \lambda_{\tau}\!\cdot\!\tau_t\Big],
\end{equation}
where $\mathrm{CE}$ is cross-entropy, $\tau_t$ aggregates sensor and network inference latency, and $(\lambda_{\text{gap}},\lambda_{\tau})$ trade-off accuracy, rate/SNR optimality, and latency. We evaluate $\mathcal{M}_\theta$ for individual modalities and their fusions to quantify accuracy, spectral-efficiency loss, SNR gap, and end-to-end latency under realistic hardware impairments.

\section{Proposed Method}\label{sec:prop_meth}
We develop a multimodal beam prediction network that estimates the receive-beam index $\hat b_t$ from synchronized ISAC observations $\mathcal{S}_t$ in the Scenario‑33 V2I setting. The design follows three principles: (i) treat the in-band communication signal as an additional sensing stream by embedding the 64‑dimensional mmWave power vector; (ii) extract modality-specific features using lightweight CNN/MLP encoders that map all sensors into a shared embedding dimension; and (iii) fuse these embeddings with a transformer-based fusion module that can flexibly operate with any subset of available modalities. The network is trained as a beam-classifier and evaluated with both accuracy and communication-centric metrics.

\subsection{Data preparation and synchronization}
Scenario‑33 provides synchronized sample per index consisting of an RGB camera frame, a 64\,$\times$\,1 mmWave receive power vector, LiDAR point cloud, radar frame, and GPS measurements. We first convert the raw logs into a compact indexed dataset. For each sample index, the preprocessing pipeline:
\begin{itemize}
    \item converts the mmWave measurements into a 64‑dimensional receive‑power feature vector;
    \item extracts LiDAR measurements as XYZ point‑cloud features and stores them in a compact representation;
    \item interprets the radar measurements as 2D radar‑intensity maps for subsequent processing;
    \item associates each sample with its corresponding camera observation; and
    \item transforms the GPS measurements into per‑sample feature vectors containing latitude, longitude, speed, and quality indicators.
\end{itemize}

All modalities are aligned by index, using the mmWave sequence as reference, and the aligned samples are split into train/validation/test using a contiguous 70/15/15 partition.



\subsection{Modality-specific Encoders}
Each sensor stream is embedded by a dedicated neural encoder that outputs a $d$‑dimensional feature vector (with a shared embedding size $d{=}256$ in our implementation). This promotes symmetry across modalities and simplifies fusion.
\subsubsection{mmWave power vector}
The 64‑dimensional mmWave receive power vector is encoded by a two-layer MLP. Concretely, the encoder applies a first linear layer from $\mathbb{R}^{64}$ to $\mathbb{R}^{2d}$ followed by a ReLU activation, and a second linear layer from $\mathbb{R}^{2d}$ to $\mathbb{R}^{d}$ followed by another ReLU. The resulting embedding
$\mathbf{f}^{(\mathbf{p}_t)}_t \in \mathbb{R}^{d}$
captures the angular power distribution across the receive codebook and serves as the communication-as-sensing feature.
\subsubsection{Camera}
For the RGB camera stream, we use a convolutional backbone that maps an image tensor $\mathbf{I}_t \in \mathbb{R}^{3 \times H \times W}$ to a $d$‑dimensional embedding. We use a ResNet‑18 model pre-trained on ImageNet and replace its classification head with a linear projection into $\mathbb{R}^{d}$. In addition, a lightweight CNN with ReLU activations is used. In both cases, the camera encoder outputs
$\mathbf{f}^{(\mathbf{I}_t)}_t \in \mathbb{R}^{d},$
which summarizes semantic and geometric context around the BS.
\subsubsection{LiDAR}
LiDAR point clouds are first converted into a bird’s-eye-view (BEV) occupancy image on a fixed $(H_{\text{bev}}, W_{\text{bev}})$ grid over a pre-defined field of view. The resulting single-channel BEV map is fed to a encoder, which shares the same structure as the CNN backbone but with $1$ input channel. This encoder applies stacked convolutions with ReLU and pooling followed by adaptive average pooling and a linear projection, yielding
$\mathbf{f}^{(L_t)}_t \in \mathbb{R}^{d}$.
This representation encodes road layout, static structures, and potential blockage geometry in a compact raster form.
\subsubsection{Radar}
Radar frames are loaded as real-valued 2D maps (after magnitude, dimensionality reduction, normalization, and resizing). These are passed to a CNN encoder, which is an architecture consisting of three convolutional layers with ReLU activations and intermediate pooling. This is followed by adaptive average pooling and a projection into $\mathbb{R}^{d}$. The radar encoder produces
$\mathbf{f}^{(\mathcal{R}_t)}_t \in \mathbb{R}^{d},$
capturing range/angle structure and motion cues that are complementary to camera and LiDAR.
\subsubsection{GPS‑RTK} The features are represented as a fixed-length vector (e.g., latitude, and longitude) which are normalized. This vector is fed to an MLP encoder identical in form to the mmWave power encoder i.e., a two-layer ReLU MLP with an intermediate width of $2d$ and output dimension $d$. The GPS branch produces
$\mathbf{f}^{(\mathbf{g}_t)}_t \in \mathbb{R}^{d},$
providing coarse geometric information (relative position and motion) that helps disambiguate beams, especially when other modalities are degraded.
\subsection{Transformer-based Multimodal Fusion}
Let $\mathcal{M}_t$ denote the set of modalities available at time $t$, and let $\mathcal{F}_t = \{\mathbf{f}^{(m)}_t : m \in \mathcal{M}_t\}$ be their embeddings. We fuse these modality embeddings using a token-based transformer encoder. First, a modality projector maps the set of embeddings into a sequence of tokens. A learnable token is prepended to the sequence, and a fixed set of learned positional embeddings is added to encode token order. This yields a tensor
where the tokens correspond to camera, LiDAR, radar, GPS, and/or mmWave embeddings, depending on which modalities are present.
We then apply a multi-layer transformer encoder with multi-head self-attention over this token sequence. 
Because the token list is constructed only from modalities present in the batch, the same architecture can operate seamlessly with any subset of sensors (e.g., mmWave-only, camera+GPS, or full fusion).
\subsection{Latency-aware evaluation}
Although latency is not explicitly part of the training loss, the implementation tracks both sensor and inference latency during evaluation. For each batch, we measure the forward-pass time of the model and aggregate it across all samples to obtain an average inference latency. This is combined with user-specified per-modality sensor latencies to estimate end-to-end latency for a given sensor combination. Alongside Top-$k$ accuracy, we compute spectral efficiency under the predicted and oracle beams, SNR gap in dB, and the gain ratio between predicted and optimal beam power. This evaluation protocol directly links multimodal perception quality to communication performance and timing constraints in the considered V2I ISAC setting.

\section{Results and Discussion}
\subsection{Evaluation Metrics}
We train and evaluate each sensor combination on DeepSense‑6G Scenario‑33 \cite{10144504}, using the receive‑beam codebook $\mathcal{B}$. Besides Top‑$k$ accuracy, we report: (i) the average SNR gap ($\mathrm{SNR\_gap}$ in dB) to the oracle beam (ii) the predicted vs.\ oracle spectral efficiencies ($\mathrm{SE}_{\text{pred}}$ and $\mathrm{SE}_{\text{opt}}$ in bps/Hz), (iii) the average rate loss ($\mathbb{E}[\Delta R]$) and (iv) the end‑to‑end latency. Hence, we present all metrics per sensor combination and compare them across combinations.


\subsection{Overall Accuracy and Learning Behavior}
\begin{figure}[htbp]
  \centering
  \includegraphics[width=0.8\linewidth]{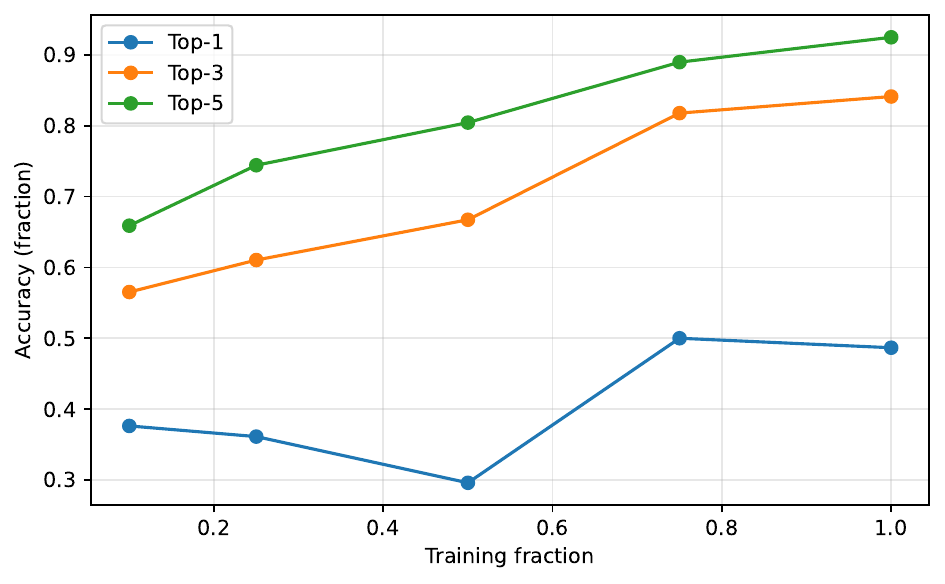}
  \caption{Learning curves for principal modalities and their fusion (Top‑$1/3/5$ accuracy vs. Dataset).}
  \label{fig:learning_curve_all}
\end{figure}

\begin{figure}[htbp]
  \centering
  \includegraphics[width=0.8\linewidth]{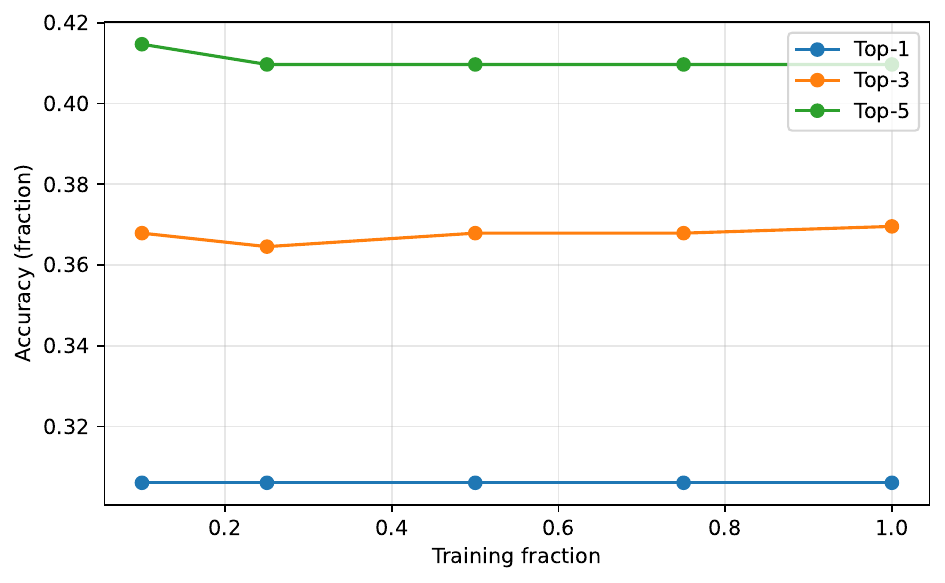}
  \caption{Learning curve for LiDAR modality.}
  \label{fig:learning_curve_l}
\end{figure}

\begin{figure}[htbp]
  \centering
  \includegraphics[width=0.8\linewidth]{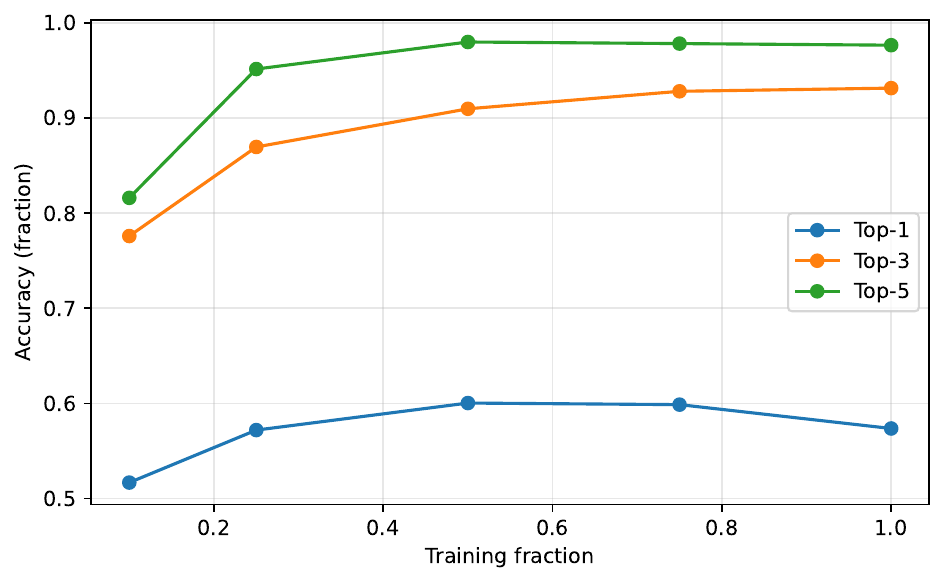}
  \caption{Learning curve for mmWave power modality.}
  \label{fig:learning_curve_m}

\end{figure}

\begin{figure}[htbp]
  \centering
  \includegraphics[width=0.8\linewidth]{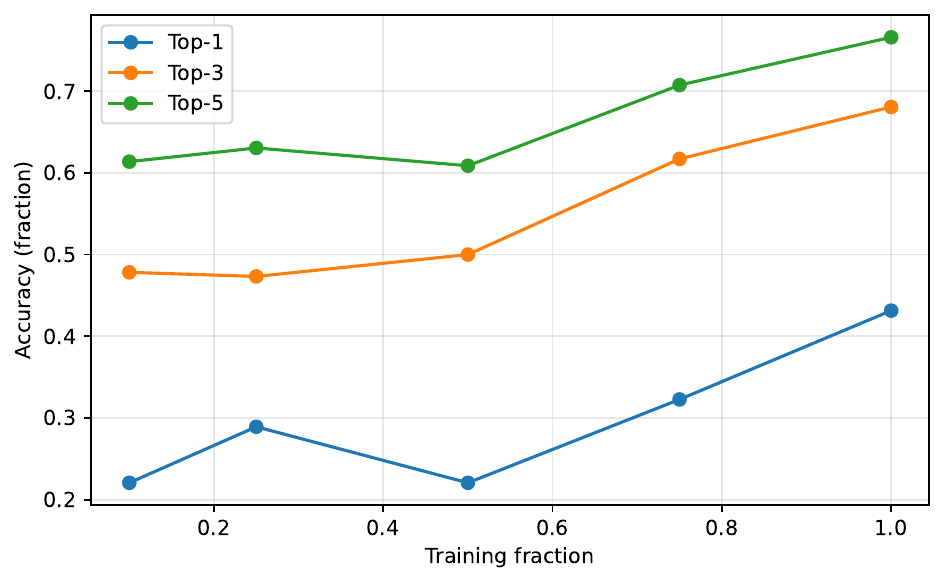}
  \caption{Learning curve for Camera modality.}
  \label{fig:learning_curve_c}
\end{figure}

\begin{figure}[htbp]
  \centering
  \includegraphics[width=0.8\linewidth]{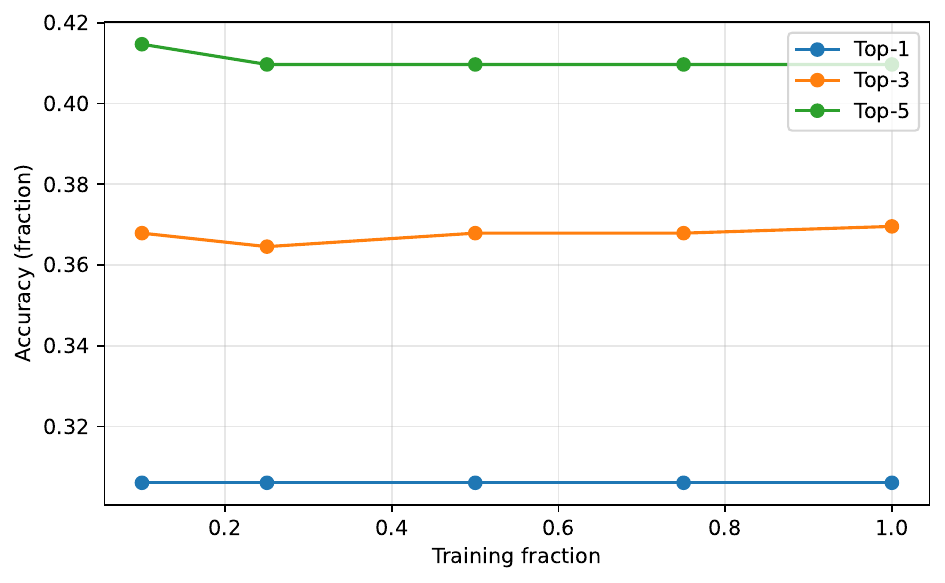}
  \caption{Learning curve for GPS modality.}
  \label{fig:learning_curve_g}
\end{figure}

\begin{figure}[!htbp]
  \centering
  \includegraphics[width=0.8\linewidth]{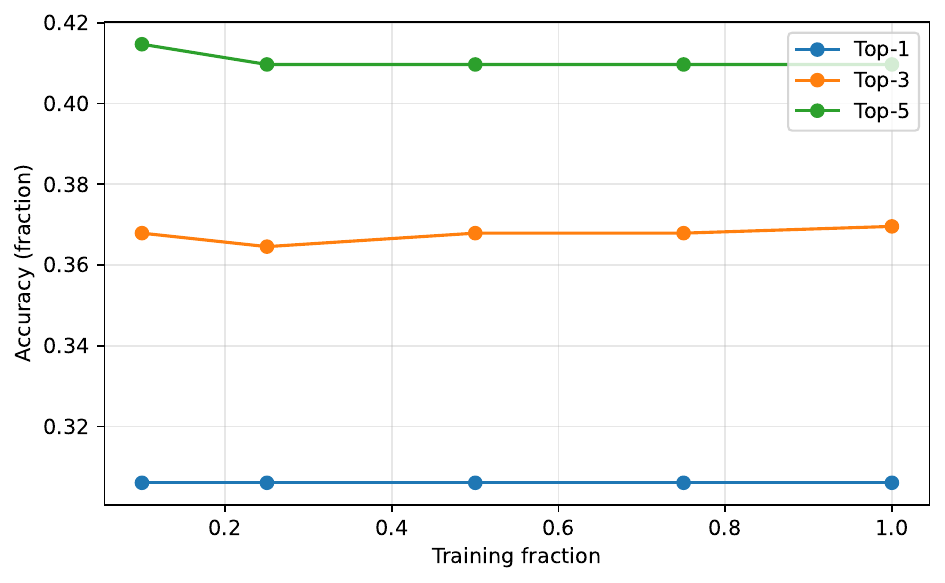}
  \caption{Learning curve for Radar modality.}
  \label{fig:learning_curve_r}
\end{figure}

Figure~\ref{fig:learning_curve_all} shows training dynamics for principal sensing modality fusion. 
The per-modality curves in Figs.~\ref{fig:learning_curve_l}-\ref{fig:learning_curve_r} underline complementary behaviors: GPS improves steadily as the model internalizes BS-UE geometry; LiDAR benefits from BEV pre‑processing, yielding rapid early gains; camera and radar progress more gradually but contribute additional improvements once fused with geometry and mmWave.

\begin{figure}[htbp]
  \centering
  \includegraphics[width=1.0\linewidth]{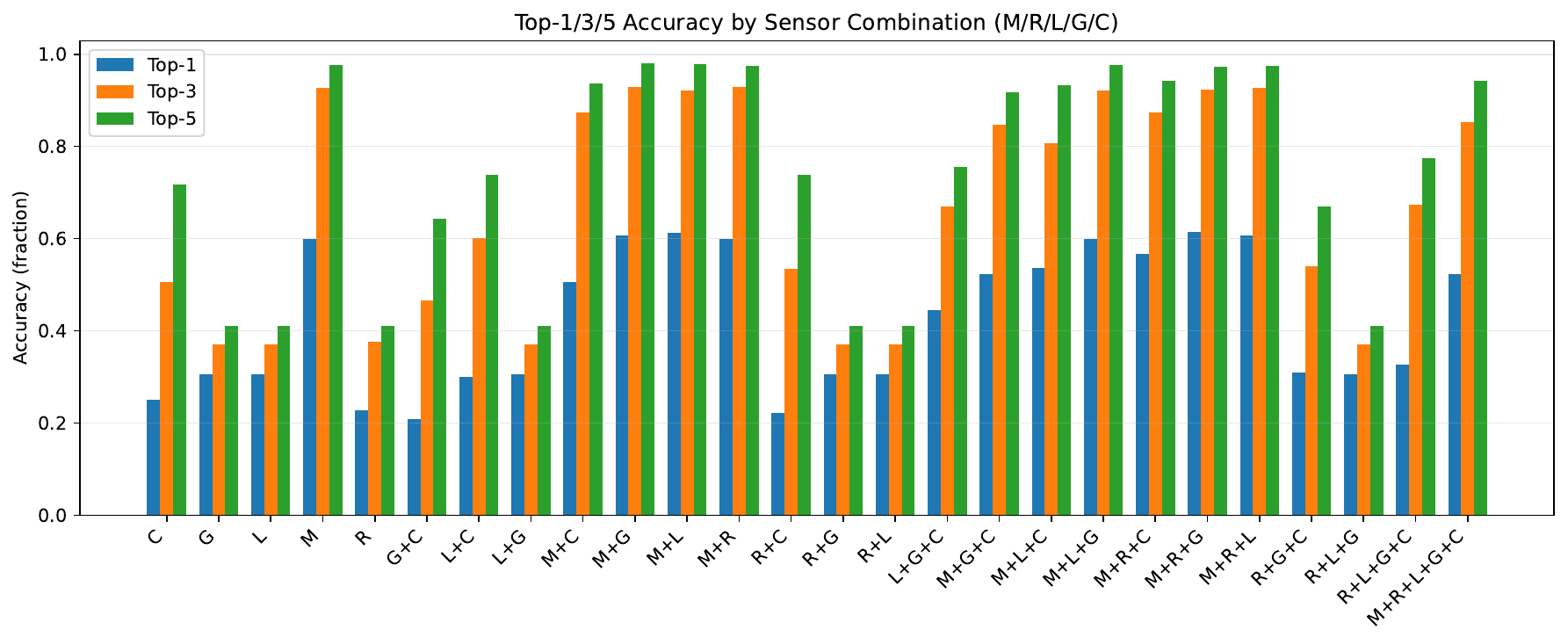}
  \caption{Top‑$1/3/5$ accuracy by sensor combination. Camera, GPS, LiDAR, mmWave and radar sensing modalities are represented with C, G, L, M and R respectively.}
  \label{fig:topk_by_combo}
\end{figure}

The cross-combination comparison in Fig.~\ref{fig:topk_by_combo} confirms that mmWave power is a strong single modality with high Top‑$3$/Top‑$5$ accuracy of approximately $93\%/98\%$. Adding GPS or radar or LiDAR consistently maintains Top‑$1/3/5$ accuracies with modest complexity as shown in Table~\ref{tab:perf_summary}.

\subsection{Spectral Efficiency and SNR Gap}
\begin{figure}[htbp]
  \centering
\includegraphics[width=1.0\linewidth]{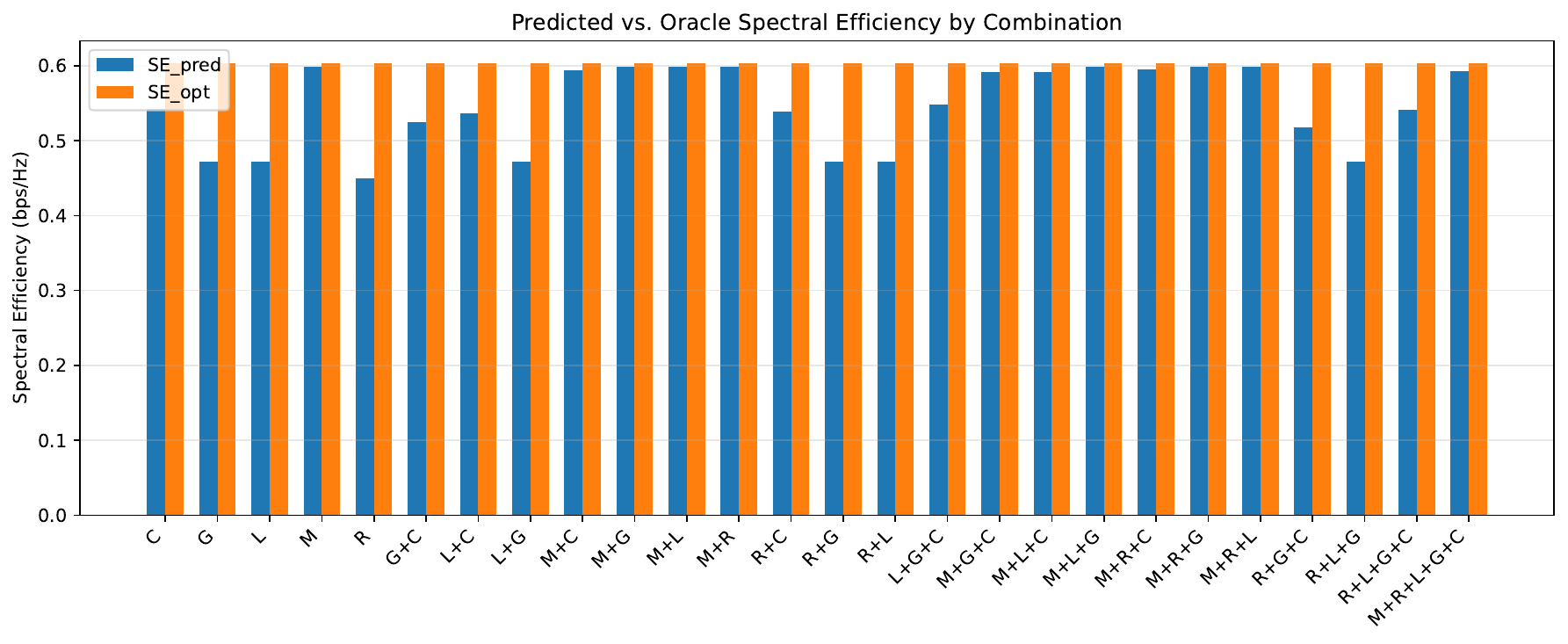}
\caption{$\mathrm{SE}_{\text{pred}}$ vs.\ $\mathrm{SE}_{\text{opt}}$ per combination. Geometry‑aided fusion (mmWave+GPS/LiDAR) narrows the SE gap disproportionately to Top‑1 changes.}
\label{fig:se_comparison}
\end{figure}

\begin{figure}[htbp]
  \centering
  \includegraphics[width=\linewidth]{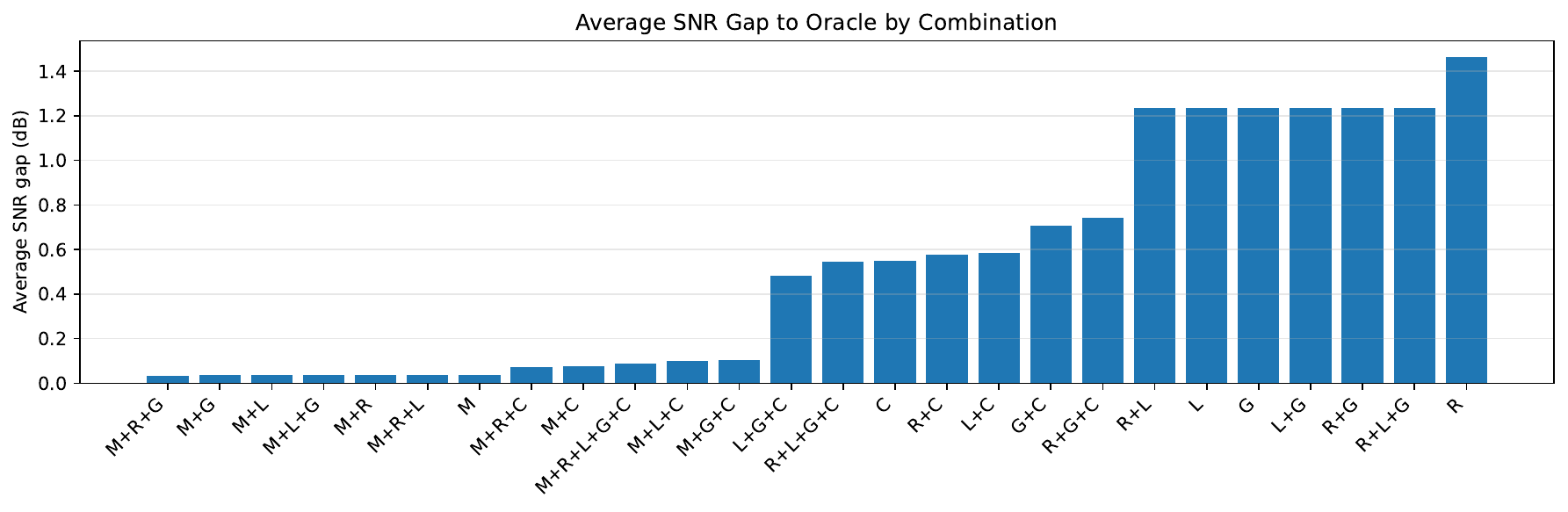}
  \caption{Average SNR gap $\mathrm{SNR\_gap}$ (in dB) to the oracle beam. mmWave‑only is sub‑dB on Scenario‑33; GPS/LiDAR further reduce the gap; camera/radar add situational gains.}
  \label{fig:snr_gap}
\end{figure}

Accuracy differences do not always translate linearly to communication quality. Figure~\ref{fig:se_comparison} compares $\mathrm{SE}_{\text{pred}}$ to the oracle $\mathrm{SE}_{\text{opt}}$. The mmWave‑only baseline already attains a small SE gap on Scenario‑33; fusing GPS (robust bearing/range cues) and LiDAR (blockage/reflector geometry) further compresses this gap, yielding disproportionately lower $\mathbb{E}[\Delta R]$ relative to the change in Top‑1 accuracy. Figure~\ref{fig:snr_gap} reports the average SNR gap as defined in Sec.~\ref{sec:sys_mod}. The mmWave‑only configuration exhibits a sub‑dB average gap, and mmWave+GPS/LiDAR reduces this further. Camera and radar contribute in low‑visibility or high‑dynamics segments, with the largest net benefit observed in fusion alongside the mmWave/GPS/LiDAR trio.

\begin{figure}[t]
  \centering
  \includegraphics[width=1.0\linewidth]{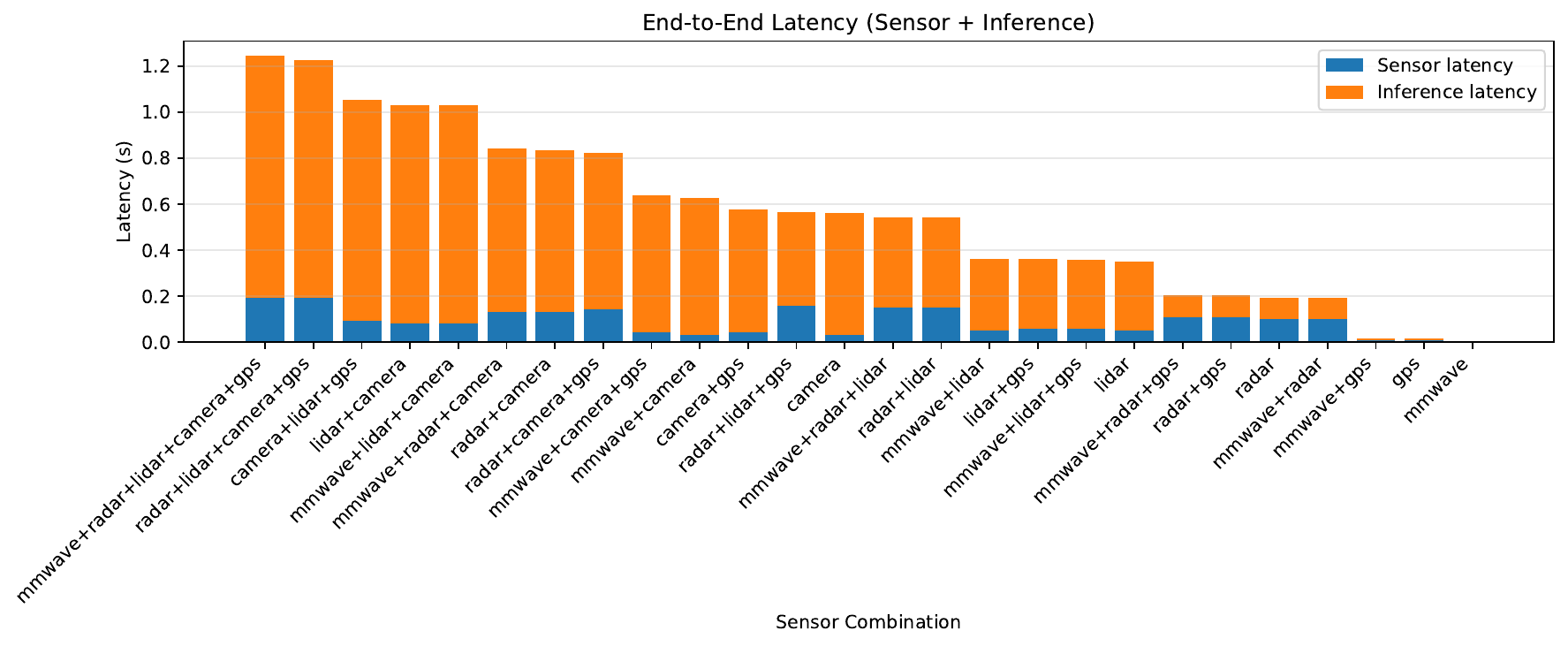}
  \caption{Decomposition of end-to-end latency into sensing and inference components for different sensor configurations.}
  \label{fig:latency_stack}
\end{figure}

\subsection{Latency-Accuracy Trade-offs}
Figure~\ref{fig:latency_stack} decomposes end‑to‑end latency into sensing and inference components. The mmWave+GPS configuration is the most latency‑efficient operating point, offering near‑baseline $\mathrm{SE}_{\text{pred}}$ at minimal additional delay. LiDAR and radar increase sensing time yet can shrink the SNR gap in mobility and intermittent‑blockage regimes. The camera branch adds semantic context at moderate overhead. Our multimodal fusion module (Sec.~\ref{sec:prop_meth}) can operate with partial modality sets, allowing costly branches (e.g., camera) to be disabled when necessary while still preserving $\mathrm{SE}_{\text{pred}}$ through mmWave+GPS.
\begin{table}[!t]
  \centering
  \caption{Performance of sensing modalities (alone and fused with mmWave) on Scenario-33.}
  \label{tab:perf_summary}
  \renewcommand{\arraystretch}{1.05}
  \setlength{\tabcolsep}{6pt}

  \sisetup{
    table-number-alignment = center,
    detect-weight          = true,
    detect-family          = true
  }

  \begin{threeparttable}
    \begin{tabular}{@{}l
                    S[table-format=1.2]
                    S[table-format=1.2]
                    S[table-format=1.2]
                    S[table-format=1.4]
                    S[table-format=1.2]
                    @{}}
      \toprule
      {\textbf{Sensing modality}} & {\textbf{Top-1}} & {\textbf{Top-3}} & {\textbf{Top-5}} &
      {\textbf{$\mathbb{E}[\Delta R]$}} & {\textbf{GR}} \\
      \midrule

      \multicolumn{6}{@{}l@{}}{\textit{Single modality}} \\
      \midrule
      mmWave   & 0.60 & 0.93 & 0.98 & 0.0044 & 0.99 \\
      GPS   & 0.31 & 0.37 & 0.41 & 0.1306 & 0.77 \\
      Radar   & 0.23 & 0.38 & 0.41 & 0.1530 & 0.72 \\
      LiDAR   & 0.31 & 0.37 & 0.41 & 0.1306 & 0.77 \\
      Camera   & 0.25 & 0.51 & 0.72 & 0.0612 & 0.89 \\
      \midrule

      \multicolumn{6}{@{}l@{}}{\textit{Fused with mmWave}} \\
      \midrule
      mmWave+GPS       & 0.61 & 0.93 & 0.98 & 0.0041 & 0.99 \\
      mmWave+Radar       & 0.60 & 0.93 & 0.98 & 0.0043 & 0.99 \\
      mmWave+LiDAR       & 0.61 & 0.92 & 0.98 & 0.0042 & 0.99 \\
      mmWave+Camera       & 0.51 & 0.87 & 0.94 & 0.0091 & 0.98 \\
      M+G+L+C+R & 0.52 & 0.85 & 0.94 & 0.0102 & 0.98 \\
      \bottomrule
    \end{tabular}

    \begin{tablenotes}[flushleft]
      \footnotesize
      \item \textbf{Abbreviations:} M = mmWave, G = GPS, L = LiDAR, C = Camera, R = Radar;
      $\mathbb{E}[\Delta R]$ = expected rate loss; GR = gain ratio.
    \end{tablenotes}
  \end{threeparttable}
\end{table}
\subsection{Data Efficiency and Ablations}
Learning-curve slices (Figs.~\ref{fig:learning_curve_l}--\ref{fig:learning_curve_r}) indicate strong data efficiency. The mmWave and GPS models saturate rapidly, whereas LiDAR, camera, and radar continue to improve with additional training data, beginning to plateau in the intermediate-data regime. In the ablation study (Table~\ref{tab:perf_summary}), we further observe that replacing a purely classification-oriented objective with the system-centric loss from Sec.~\ref{sec:sys_mod}, which explicitly couples the learning signal to rate and SNR consistently lowers both $\mathrm{Rate\_loss}$ and $\mathrm{SNR\_gap}$ while preserving Top-$k$ accuracy. This outcome better aligns the predictor with the system-level objectives defined earlier.

\FloatBarrier
\section{Conclusion}

We presented a system-centric comparative analysis of ISAC beam prediction on DeepSense-6G Scenario-33. Our latency-aware fusion model treats the mmWave power vector as a sensing feature alongside camera, LiDAR, radar, and GPS, and is trained using a rate-aware loss. The results indicate that mmWave provides a strong anchor signal; GPS and LiDAR deliver consistent gains; and camera and radar offer situational improvements. Notably, even small changes in Top-1 accuracy can translate into meaningful reductions in spectral-efficiency and SNR gaps. Operationally, we recommend defaulting to mmWave+GPS and selectively enabling additional sensors under hardware constraints. Overall, the framework serves as a reproducible baseline that respects practical limitations and supports cross-scenario, temporal, and near-field extensions.
 

\section*{Acknowledgment}
This work was jointly supported by the African Center of Excellence in Internet of Things (ACEIoT) University of Rwanda,
Regional Scholarship and Innovation Fund (RSIF), and National Research Foundation of Korea under Grant RS-2024-00409492.
\bibliographystyle{IEEEtran}
\bibliography{IEEEabrv,References}

\end{document}